\documentclass[twocolumn]{article}

\usepackage{amsmath}
\usepackage{graphicx}
\usepackage[affil-it]{authblk}
\usepackage[colorlinks,citecolor=blue,breaklinks=true]{hyperref}

\newcommand{\eqnRef}[1]{eq.~\ref{eqn:#1}}
\newcommand{\eqnRefs}[2]{eqs.~\ref{eqn:#1} and \ref{eqn:#2}}
\newcommand{\figRef}[1]{fig.~\ref{fig:#1}} 
\newcommand{\figRefs}[2]{figs.~\ref{fig:#1} and~\ref{fig:#2}} 

\newcommand\vopar        {\boldsymbol{M}}
\newcommand\opar         {M}
\newcommand\TOP          {\mathcal T}
\newcommand\Tgl          {\TOP_{gl}}
\newcommand\TglLT        {\TOP^{LT}_{gl}}
\newcommand\TglTD        {\TOP^{2D}_{gl}}
\renewcommand\phi        {\varphi}
\newcommand{\entrop}      {{\mathbf H}}
\newcommand\bTheta       {\boldsymbol\Theta}
\newcommand{\tlogtwo}{\log_2}
\providecommand{\e}[1]{\ensuremath{\times 10^{#1}}}
\newcommand{\ie}[0]      {\textit{i.e.}}
\providecommand{\etal}[0]    {\textit{et al.}}

\newcommand{\nicesim}{{\raise.17ex\hbox{$\scriptstyle\mathtt{\sim}$}}}

\title{Information flow in finite flocks}

\author{J. Brown \thanks{Email: \texttt{jbrown@csu.edu.au}; Corresponding author}} \affil{School of Computing \& Mathematics, Charles Sturt University, Bathurst, NSW, Australia}
\author{T. Bossomaier} \affil{Centre for Research in Complex Systems, Charles Sturt University, Bathurst, NSW, Australia}
\author{L. Barnett} \affil{Sackler Centre for Consciousness Science, Department of Informatics, University of Sussex, Brighton, U.K.}

\begin{document}

\maketitle

\begin{abstract}

We simulate the canonical Vicsek model and estimate the flow of information as a
function of noise (the  variability in the extent to which each animal aligns with
its neighbours). We show that the global transfer
entropy for finite flocks not only fails to peak near
the phase transition, as demonstrated for the canonical 2D Ising model, but remains constant from the transition to very low noise
values. This provides a foundation for future study regarding information flow in more complex models and real-world flocking data.
\end{abstract}

\section{\label{sec:intro}Introduction}

Recent experimental studies of animal flocks,
fish~\cite{calovi15,rosenthal15}, birds such as pigeons~\cite{nagy13},
starlings~\cite{cavagna10:starlings}, midges~\cite{attanasi14:midges} and
sheep~\cite{ginelli15} have dramatically increased our understanding
of flocking dynamics. A central  theoretical issue is how the
communication range
between flock members leads to global coordination of the flock,
leading to proposals~\cite{mora11,attanasi14:midges} that the flock exists at some critical point---the point at which flocks transition between order and disorder---providing the ideal scenario for collective reaction to external stimuli, with enough order to form collective behaviour without overwhelming inertia. This carries concomitant implications for information flow.

We measure information flow among members of these flocks, using transfer entropy~\cite{schreiber00}, in particular Global Transfer Entropy (GTE, \eqnRef{Tgl}), shown to peak on the disordered side of the phase transition in the  2D Ising spin model~\cite{barnett13}, where mutual information (MI) and pairwise transfer entropy (TE) peak \emph{at} the transition.

Real world flocks are
\emph{active matter}---systems far from equilibrium, which do not
conserve momentum or other dynamical quantities~\cite{fodor16}.
We analyse the long-term limit GTE of the minimalist  Standard Vicsek
Model
(SVM)~\cite{vicsek95} of
collective motion~\cite{Toner98,vicsek12}  using a novel
closed-form dimensional reduction, obtained by exploiting an
approximate isometry in the SVM. This approach has demonstrated anomalous behaviour in MI~\cite{Barnett18}, diverging as noise tends to zero and failing to peak
at the transition.

This study reveals two unexpected information-theoretic differences
from  the Ising model~\cite{barnett13} at long observation times: \emph{absence of a peak in GTE on the disordered side of the phase
  transition; and a slowly changing value below
  the transition}\footnote{The disordered side is the high noise side of
  the transition and is in this sense ``above'' the transition.}.
For shorter observation times, unlike the Ising model, the GTE peaks \emph{at}, rather than above, the phase transition.

While the final closed-form expression for GTE (\eqnRef{TglLT})
requires an isometry approximation,  two technical
innovations (\eqnRefs{Tgl3D}{Tgl2D})
were needed to  reduce computational requirements (without isometry):
replacing the multi-dimensional vector of interacting particles with a
consensus vector; and exploiting the independence of the noise,
leading to  the surprising result (\eqnRef{Tgl2D}) that
calculation of global information flow requires \emph{no} measurement
of neighbouring particles.

This new behaviour may inform future studies on real-world flocks or other more sophisticated models, noting that for the SVM maximum information flow occurs not just near the phase transition, but throughout the entire ordered regime as well.

\section{The Standard Vicsek Model}
The SVM comprises a set of $N$ point particles (labelled $i = 1,\ldots,N$) moving on a plane of linear extent $L$ with periodic boundary conditions (see~\cite{GTESuppMat}~for full details). Each particle moves with constant speed $v$, and interacts only with neighbouring particles within a fixed radius $r$. Positions $\vec{x}_i(t)$ and headings $\theta_i(t)$ are updated synchronously at discrete time intervals $\Delta t = 1$ according to

\begin{align}
	\vec{x}_i(t + \Delta t) &= \vec{x}_i(t) + \vec{v}_i(t)\Delta t \label{eqn:OVAposUpdate} \\
	\theta_i(t+\Delta t) &= \varphi_i(t) + \omega_i(t) \label{eqn:OVAvelUpdate}
\end{align}

respectively, $\varphi_i(t)$ is the average heading of all particles
within the interaction radius, $r_i$, of particle $i$ (including particle $i$
itself), and $\omega_i(t)$ is white noise uniform on the interval
$[-\eta/2, \eta/2]$ with intensity $\eta \in (0,2\pi]$. The average heading~\cite{vicsek95}, which constructs the consensus vector, is $\varphi_i(t)=\arctan [ \langle\sin(\theta(t))\rangle_{i,r} / \langle\cos(\theta(t))\rangle_{i,r}]$, where $\langle z\rangle_{i,r}\equiv\frac1N\sum_j^Nz_j\delta_r(i,j)$ and $\delta_r(i,j)$ is $1$ if neighbour $j$ is within the interaction radius, $r$, of particle $i$, and $0$ otherwise; note that $\delta_r(i,i)=1$, so that $z_i$ is always included in the average. The velocity vector $\vec{v}_i(t)$ is constructed from the heading $\theta_i(t)$ with constant speed $v$.
Particle density $\rho = N/L^2$ is fixed throughout at $0.25$.

Considering the SVM as a steady-state \emph{statistical ensemble} of
finite size containing $N$ particles with control parameter $\eta$, we use capitals to
indicate quantities sampled from the ensemble; in particular,
$\Theta_i$ denotes an ensemble sample of the heading of the $i$th
particle. The full \emph{order parameter} for the SVM ensemble is the
2D mean particle velocity vector $\vopar$ with magnitude $\opar \in
[0,1]$ and heading $\Phi \in (0,2\pi]$~\cite{GTESuppMat}. $M = 1$
  iff all particles are aligned, while in the
  fully-disordered case ($\eta = 2\pi$) we have $\opar \to 0$ in the
  large-system limit $N \to \infty$. The ensemble variance

\begin{equation}
 \chi = \langle \opar^2\rangle-\langle \opar\rangle^2
\label{eqn:suscep}
\end{equation}

defines the \emph{susceptibility}; a peak in $\chi$ as a function of $\eta$ is taken to locate an (approximate) phase transition~\cite{baglietto08:finite}\footnote{Technically, this assumes the fluctuation-dissipation theorem~\cite{Landau00}, which the SVM does not obey; however, the quantity is widely used in studies of the SVM~\cite{baglietto08:finite,Aldana09,Chat2008Collective}.}.

The phase transition in the original SVM was thought to be second order, but this was disputed~\cite{gregoire04,Chat2008Collective} and it transpired that seemingly minor details affect the nature of the transition:
type of noise statistics~\cite{chepizhko09}; forward versus backward{}
updating (especially at high particle velocities)~\cite{nagy07};
boundary conditions associated with density bands or spin
waves~\cite{aldana07}; and the cone of influence on each
particle~\cite{durve16,romensky14}. The consensus for SVM now appears to state that the phase transition is second order for low velocities, and first order for high velocities~\cite{Aldana09,baglietto12,Bahar18}. Consequently, we utilise the original SVM model (backward updating, angular noise, periodic boundary conditions and low density) over a range of velocities. Observation of the Binder cumulant~\cite{Binder87} for these regimes (not shown) indeed shows a sharp minimum---representative of a first order transition---only at high velocity magnitudes ($v=2.00$), consistent with~\cite{Aldana09}.

The finite-size SVM exhibits
behaviours~\cite{Barnett17} akin to ``continuously-broken ergodicity''~\cite{mauro07}.
Over short observation windows the SVM is confined to a comparatively small volume of
phase space, thus breaking symmetry and ergodicity. As the  window
increases, the SVM explores progressively larger volumes of phase
space, until ergodicity is restored, albeit requiring very long windows at
low noise.

Thus  the ensemble statistics are  observation
time-dependent~\cite{Barnett17}, giving two
regimes---\emph{short-term} and \emph{long-term} statistics. In the
short-term regime, we collate statistics (with no ergodic assumptions) over ranges of observation sizes spanning several orders
of magnitude, demonstrating the effect of observation time. In the long-term limit,
since ergodicity is unbroken, the time average of the GTE will be equal to the ensemble average, and thus statistics can be measured from ensembles constructed of many independent realisations---each with shorter observation windows---rather than the prohibitively long time spans required for solitary realisations to traverse the entire phase space.

\section{Global Transfer Entropy}
The information flow between two continuous random processes, from $Y$ to $X$, is given by the TE~\cite{schreiber00}~:

\begin{equation}
\TOP_{Y \to X} = \entrop(X' \,|\, X) - \entrop(X' \,|\, X, Y)\,,
\label{eqn:te}
\end{equation}

where $X,Y$ are the process histories, and $X'$ is the updated state. We truncate process histories to just the most recent state as in~\cite{barnett13}. For a continuous random variable $X$, $\entrop(X) \equiv -\int p_X(x)\log p_X(x) dx$ denotes \emph{differential entropy}, which necessitates a continuous estimator~\cite{kraskov04}. GTE extends TE to measure information flow from the entire system to a single particle, and here is defined as the ensemble statistic

\begin{equation}
	\Tgl \equiv \TOP_{\bTheta \to \Theta_I} =  \entrop(\Theta'_I \,|\, \Theta_I) - \entrop(\Theta'_I \,|\, \bTheta) \label{eqn:Tgl}
\end{equation}

where $I$ is uniform on
the set of particle indices, $\Theta'_i$ is the updated heading at
time $t+1$ of the $i$th particle, $\bTheta = (\Theta_1,\ldots,\Theta_N)$ is the vector of all $N$
particle headings.

Since the update of a particle's heading is
mediated purely by the consensus heading of its neighbours, the GTE
may be reduced to three dimensions,  \ie, $\entrop(\Theta'_I\,|\, \bTheta) = \entrop(\Theta'_I\,|\,
\Phi_I)$, giving:

\begin{equation}
\Tgl \equiv \TOP_{\Phi_I \to \Theta_I} =  \entrop(\Theta'_I \,|\, \Theta_I) - \entrop(\Theta'_I \,|\, \Phi_I)\,, \label{eqn:Tgl3D}
\end{equation}

thus eliminating dimensionality issues surrounding $\bTheta$.

Noting that $\theta'_i = [\varphi_i + \omega_i]$ for any
particle $i$---where $[\ldots]$ denotes modulo $2\pi$,
confining the result to $(-\pi,\pi]$---and that noise $\omega_i$
  is independent of  $\varphi_i$ we have just $\entrop(\Theta'_I\,|\,
  \Phi_I) = \entrop(\Omega)$ where $\Omega$ is the (ensemble) noise.

Thus \emph{all measurement of particle neighbours is eliminated} and \eqnRef{Tgl3D} reduces to the two dimensional:

\begin{equation}
	\TglTD  =  \entrop(\Theta'_I \,|\, \Theta_I) - \entrop(\Omega)\,. \label{eqn:Tgl2D}
\end{equation}

Finally, in the long term limit rotational symmetry remains approximately unbroken: that is, for
any fixed angle $\alpha$, the joint distribution
$(\Theta_1+\alpha,\ldots,\Theta_N+\alpha)$ is the same as the joint
distribution $(\Theta_1,\ldots,\Theta_N)$. Under this \emph{isotropy} approximation (see below), \eqnRef{Tgl2D} reduces
to a one-dimensional form in which only changes in particle heading
$\theta'_i-\theta_i$ and noise $\Omega$ appear. Let
$p(\theta_1,\theta_2)$ be the probability density function (pdf) of
$(\Theta'_I,\Theta_I)$~\cite{GTESuppMat}. Under the assumption of
rotational symmetry we have:

\begin{equation}
p(\theta_1,\theta_2) = \frac{1}{2\pi}q(\theta_1-\theta_2)\,,
\label{eqn:oneDimPdf}
\end{equation}

where $q(\theta)$ is the pdf of $\Theta_I-\Theta'_I$. Since  the marginal distributions of $\Theta_I$ and $\Theta'_I$ are uniform on the unit circle in the long-term (ergodic) limit, we obtain $\entrop(\Theta'_I\,|\, \Theta_I) = \entrop([\Theta'_I-\Theta_I])$ which reduces \eqnRef{Tgl2D} to the closed-form expression

\begin{equation}
  \TglLT = \entrop([\Theta'_I - \Theta_I]) - \entrop(\Omega)\,,
  \label{eqn:TglLT}
\end{equation}

for the long-term GTE, where $[\cdots]$ denotes the internal angle. Note that at $\eta=2\pi$, $\entrop([\Theta'_I - \Theta_I]) = \entrop(\Omega) = \log2\pi$  and thus $\TglLT$ vanishes at maximum noise, as expected. As noise decreases, the particles align more and more strongly, so that the distributions of both $\Theta'_I - \Theta_I$ and $\Omega$ become increasingly sharply peaked. Since these are both \emph{differential} entropies, they both diverge to $-\infty$. The exact nature of the divergence is established in simulations discussed below.

The isotropy approximation arises because the  SVM on a 2D plane with periodic boundary conditions---\ie~a
flat torus---is not in fact strictly isotropic. We tested its validity by repeating the long-term
simulations while randomly rotating the SVM frame of reference between
each update, thus enforcing isotropy~\cite{baglietto09}.
Negligible error was introduced around the phase transition and
at very low noise~\cite{GTESuppMat}.

\section{Results And Discussion}

\begin{figure}
\includegraphics[width=\columnwidth]{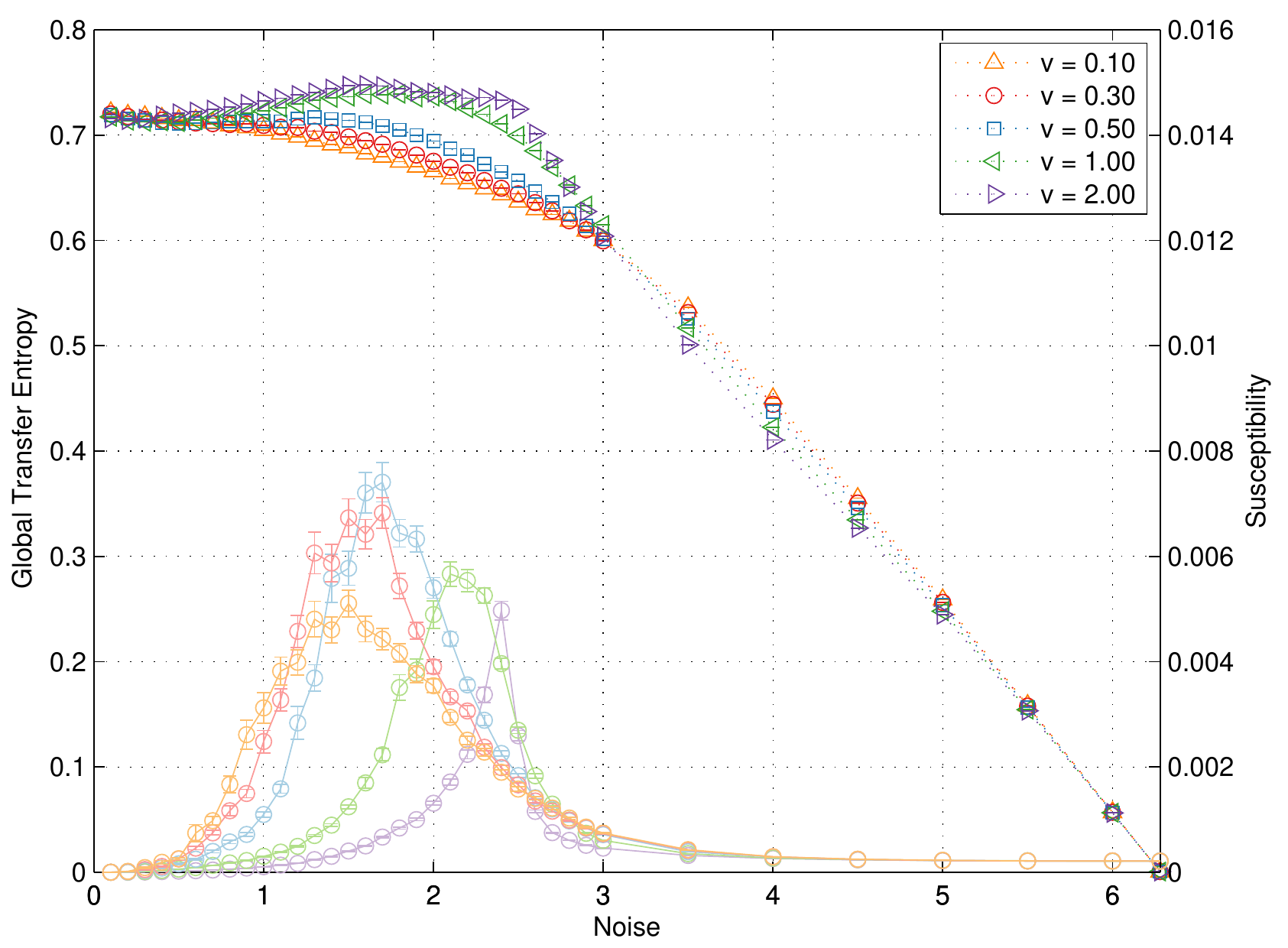}
\caption{Long-term GTE $\TglLT$ (dotted) calculated according to \eqnRef{TglLT} for a range of particle velocities for $0<\eta\leq2\pi$. System size $N=1000$ particles, density $\rho=0.25$ and velocities $v$ as indicated. Simulation: ensembles constructed from $20$ realisations at observation time $T=500$ time steps each, under ergodic assumptions and the isotropy approximation (see text). Error bars at $1$ s.e. (smaller than symbols) were constructed by $10$ repetitions of the experiment. Lines show susceptibility $\chi$~(\eqnRef{suscep}).} \label{fig:TglLT}
\end{figure}

Figure~\ref{fig:TglLT} shows the long-term GTE $\TglLT$ estimated in
sample according to \eqnRef{TglLT} for a range of particle
velocities. For $v < 0.5$ there is no peak in
GTE and for $v \geq 0.5$ peaks occur at or below the phase transition---identified as a peak in $\chi$ as per \eqnRef{suscep}---with all GTE values approaching approximately $0.72$ bits as noise tends to zero. As $v$ increases, a shift in the $\chi$ peak to higher noise values occurs.

\begin{figure}
\includegraphics[width=\columnwidth]{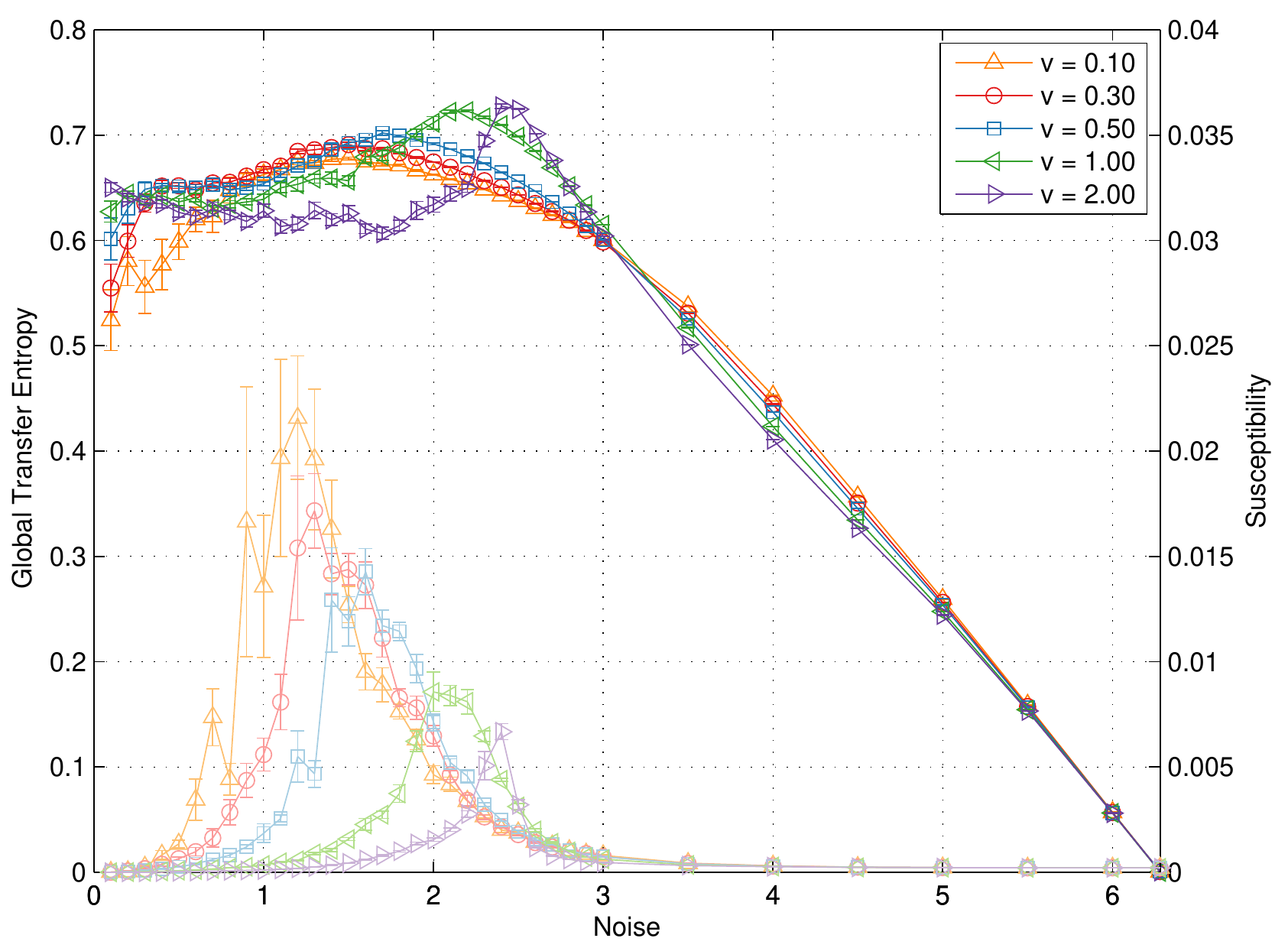}
\caption{$\TglTD$ (dark lines) and susceptibility $\chi$ (light lines) estimated for a range of particle velocities (parameters as in \figRef{TglLT}). $\TglTD$ was estimated according to \eqnRef{Tgl2D} over $T=5000$ time steps (that is, with no ergodic assumption) after relaxation to steady state, using a nearest-neighbour estimator (full simulation details in~\cite{GTESuppMat}). $\chi$ was estimated over the same realisations. Error bars at $1$ s.e. constructed by $10$ repetitions.} \label{fig:gte-3d}
\end{figure}

For \emph{short observation times,} by contrast, $\Tgl$ and $\TglTD$ estimated
according to \eqnRefs{Tgl3D}{Tgl2D} respectively (and with no isotropy assumption) do peak at the transition, rather than in the disordered regime as in the
Ising model~\cite{barnett13}; see \figRefs{gte-3d}{gte-2d}.
\begin{figure}
\includegraphics[width=\columnwidth]{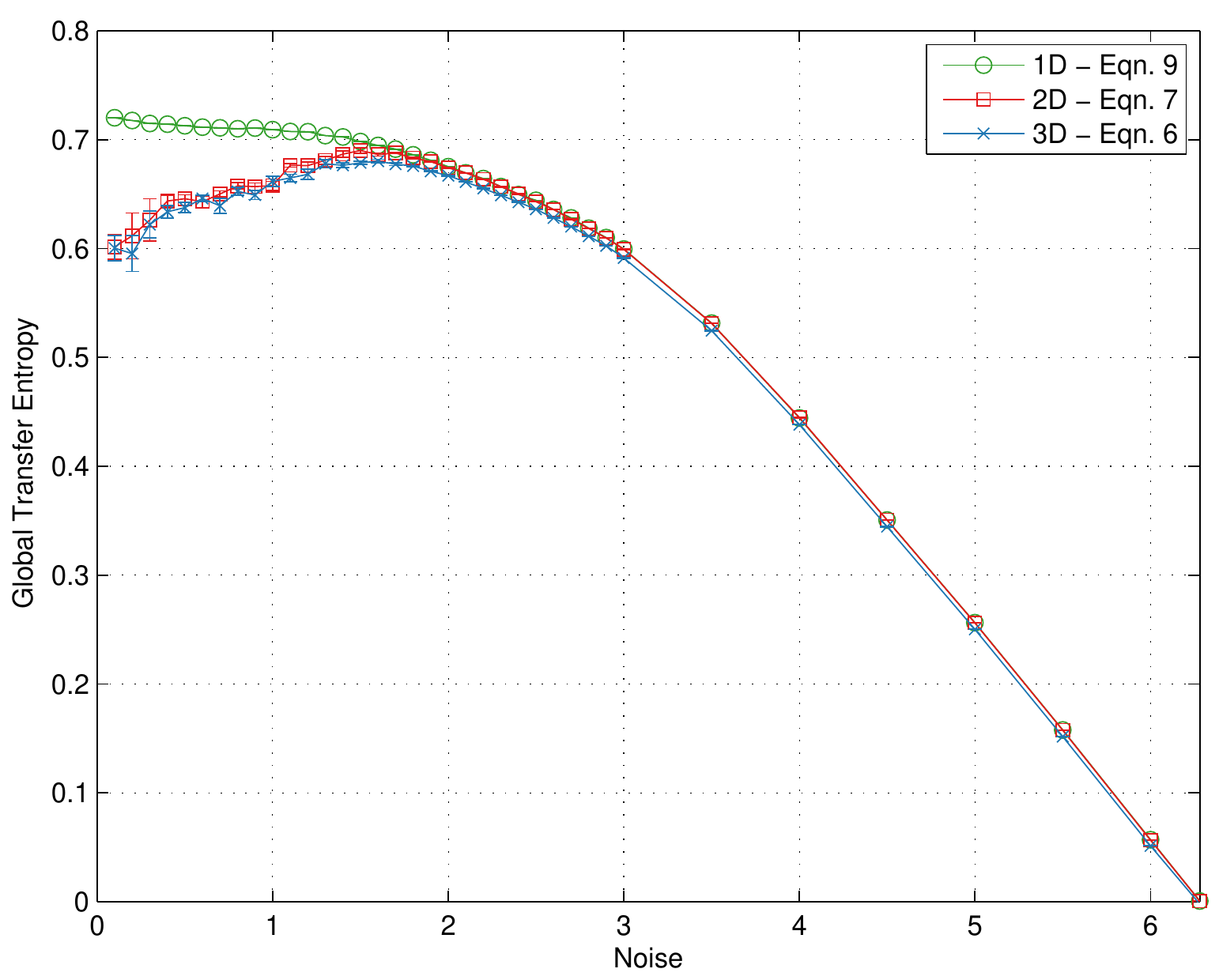}
\caption{$\Tgl$ as measured by the one-dimensional form (\eqnRef{TglLT}, circles), the three-dimensional form (\eqnRef{Tgl3D}, crosses) and the two-dimensional form (\eqnRef{Tgl2D}, squares) for $v=0.30$. Error bars at $1$ s.e. constructed by $10$ repetitions.} \label{fig:gte-2d}
\end{figure}

Figure~\ref{fig:gte-2d} shows the effect of eliminating the consensus vector measurement in \eqnRefs{Tgl3D}{Tgl2D}. The agreement between the two is extremely close, although \eqnRef{Tgl2D} gives slightly better results for numerical reasons.

\begin{figure}
\includegraphics[width=\columnwidth]{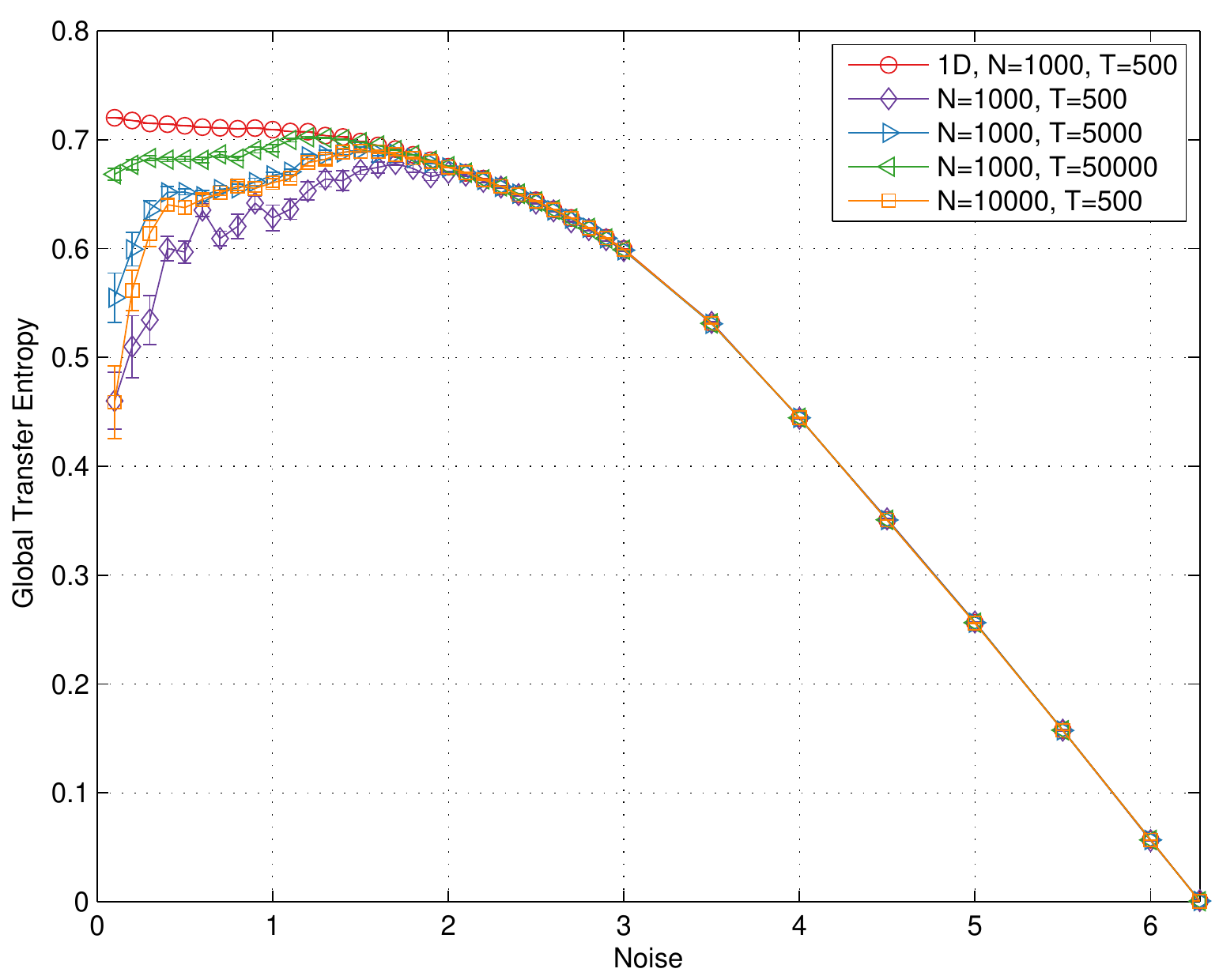}
\caption{$\TglTD$ estimates according to \eqnRef{Tgl3D} at fixed velocity $v = 0.30$ for a range of observation times $T$ as indicated, along with the long-term $\TglLT$ of \eqnRef{TglLT} as per \figRef{TglLT}. System sizes $N$ as indicated, other simulation details as for previous figures. Note that the blue right triangle line ($N=1000, T=5000$) and orange square line ($N=10000, T=500$) lie on top of each other and represent constant $NT$.}
\label{fig:gte-window-size}
\end{figure}

Some flattening at low noise occurs, particularly for higher velocities. Here GTE does not converge to the $\nicesim0.72$ bits observed in the $\TglLT$. The shorter the observation window, the nearer we are to ergodicity-breaking as in the Ising Model~\cite{barnett13}~and thus GTE $\to 0$ as $\eta \to 0$. This is confirmed in \figRef{gte-window-size} which shows $\TglTD$ for a single fixed velocity at observation window size varying over two orders of magnitude, along with the long-observation time limit $\TglLT$. As observation time increases, the GTE peak flattens and constant GTE in the ordered regime starts to occur, approaching, as predicted, the long-observation time limit.

\begin{figure}
\includegraphics[width=\columnwidth]{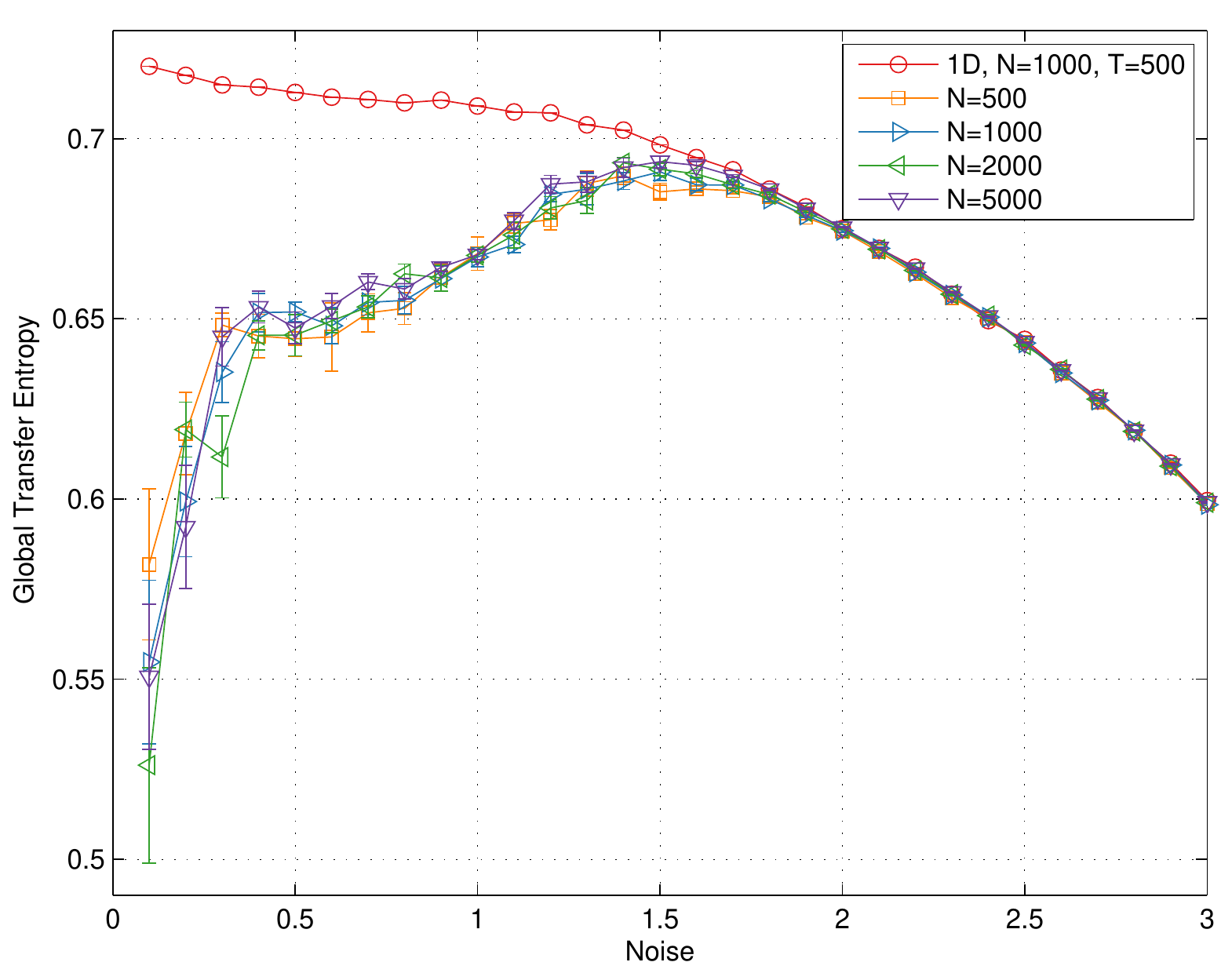}
\caption{$\TglTD$ estimates according to \eqnRef{Tgl3D} at fixed velocity $v = 0.30$ for $T=5000$ time steps for varying $N$, along with $\TglLT$ according to \eqnRef{TglLT} (red circles).}
\label{fig:nScaling}
\end{figure}

Finally, \figRef{nScaling} shows the effect of varying the system size. For $\eta > 0.4$, $\TglTD$ increases---converging to $\TglLT$---as $N$ increases. Below this however, $\TglTD$ diverges further as $N$ increases, reflecting the reduced capacity of the system to explore large volumes of the phase space. Near the transition, flocks are in flux, breaking apart and reforming. Given this fluidity, a larger number of particles results in more sub-flocks, which consequently are able to explore the phase space more efficiently, so that the GTE approaches $\TglLT$. At near-zero noise levels, however, flock stability predominates, with phase space exploration effected mostly by the flock's random walk-like behaviour\footnote{Although it is still possible for flocks to break apart over time.}~\cite{Brown17a}. The magnitude of the random walk is inversely proportional to the number of interacting particles; \ie, as the number of interacting particles increases, the mean of the consensus heading at $t+\Delta t$ more closely matches the mean at $t$. Due to the slower random walk, the system explores less of the phase space---more closely approximating ergodicity-breaking---and hence diverges from the long term limit $\TglLT$.

Simulation establishes the nature of the diverging entropies, showing convergence to $\nicesim0.72$ bits as $\eta \to 0$. Analysis of how particle headings evolve over time~\cite{GTESuppMat}~reveals an approximate Gaussian distribution\footnote{Strictly speaking it is a Von Mises (circular) distribution; however, as the width of the distribution is much less than $2\pi$ at and below the transition, we may safely adopt a Gaussian approximation. See~\cite{GTESuppMat}~for more details.} of heading differences---\ie, $\Delta\Theta$---as well as an approximately Gaussian distribution in the heading of the consensus vector---relative to the appropriate particle---as noise tends to zero. From the heading update in \eqnRef{OVAvelUpdate}---with particular note of the definition of $\TglLT$ in \eqnRef{TglLT}---we have:

\begin{align}
\theta_i(t+\Delta t) &= \varphi_i(t)+\omega_i(t)\,,\tag{\ref{eqn:OVAvelUpdate} revisited}\\
\theta_i(t+\Delta t) - \theta_i(t) &= \varphi_i(t)+\omega_i(t) - \theta_i(t)\,, \label{eqn:deltaThetaExpanded}
\end{align}

which allows us to decompose $\Delta\Theta$ into two independent distributions, defined by $\varphi_i(t) - \theta_i(t)$ and $\omega_i(t)$. The relative consensus heading, $\varphi_i(t) - \theta_i(t)$, is approximately Gaussian with support approximately equal to $[-\frac\eta2,\frac\eta2]$. By definition, noise $\omega_i(t)$ is uniform with support $[-\frac\eta2,\frac\eta2]$. By the Central Limit Theorem~\cite{Billingsley95}, summing these two distributions as per the RHS of \eqnRef{deltaThetaExpanded}, yields a truncated Gaussian with range $[-\eta,\eta]$ and variance twice that of the noise; \ie, $\sigma^2_{\Delta\Theta}=2\sigma^2_\Omega$. Empirical results match this, with $\sigma^2_{\Delta\Theta} = c\sigma^2_\Omega$ where $c\to2^-$ as $\eta\to0$.

Thus, closed form entropies for Gaussian and uniform distributions can be substituted into \eqnRef{TglLT}:

\begin{equation}
\begin{split}
\TglLT &= \frac12 \tlogtwo2\pi e \sigma_{\Delta\Theta}^2 - \tlogtwo\sqrt{12\sigma_\Omega^2}\,,\\
&= \frac12 \tlogtwo\frac{2\pi e c\sigma_\Omega^2}{12\sigma_\Omega^2}\,,\\
&= \frac12 \tlogtwo\frac{\pi e c}{6}\,,
\end{split}
\label{eq:tgl_final}
\end{equation}

which tends to $0.7546^-$ bits as $c\to2^-$, in reasonable agreement with simulation results, given the approximations involved.

Above the phase transition however, the distribution of $\Theta'_I - \Theta_I$ is no longer approximately Gaussian in nature. As $\eta\to2\pi$, $\Delta\Theta$ becomes increasingly convolved with a uniform distribution before reaching uniformity at $\eta=2\pi$, leading to the decrease seen in $\Tgl$ over $\eta_c < \eta < 2\pi$.

\begin{figure*}
\begin{center}
\resizebox{0.99\textwidth}{!}{\includegraphics{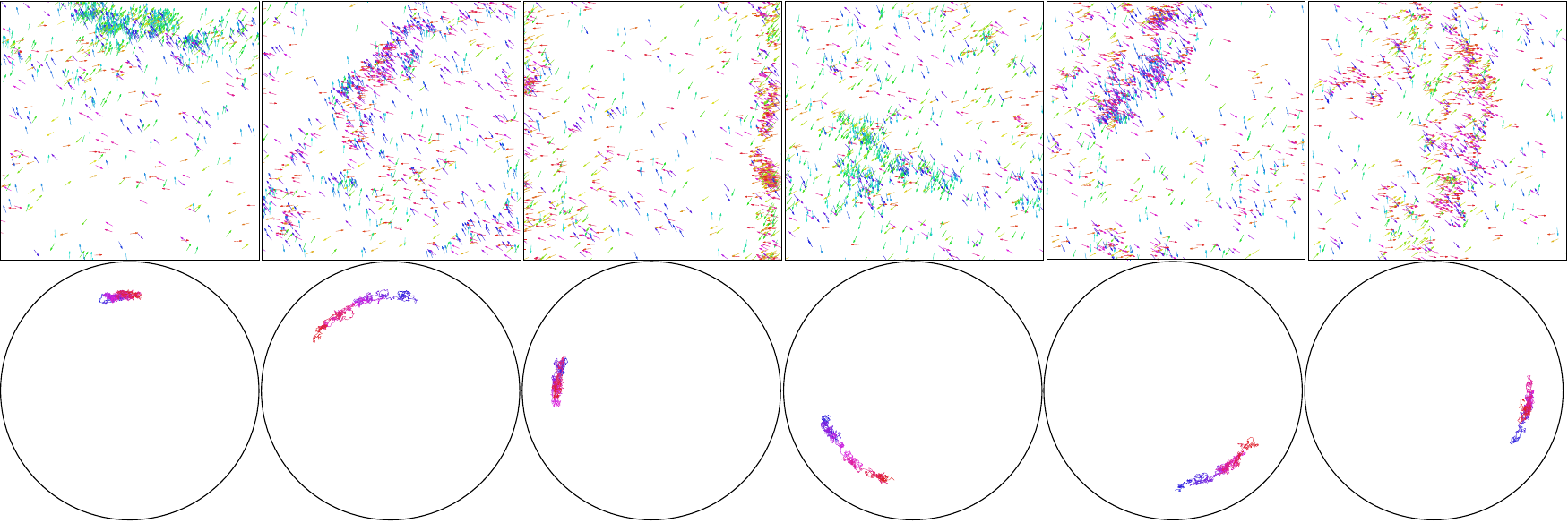}}
\caption{Snapshots from a single simulation demonstrating random walk of heading of high density bands of a flock with $N=1000$ particles at high velocity ($v=2.0$). Snapshots taken at, from left to right, $t=23\e3, 24\e3, 28\e3, 40\e3, 47\e3, 49\e3$. Top row shows the state of the flock, while the bottom row shows the two-dimensional order parameter $\vopar$---that is, mean particle velocity---for the previous $1000$ time steps going from blue ($t-1000$) to red ($t$). Distance from the center of the circle corresponds to the order parameter magnitude $\opar = |\vopar|$. Note that, as witnessed by the first two snapshots, the change in heading can be rapid, with only $1000$ time steps required for the band to precess $\pi/4$ radians. Reprinted from \cite{Barnett17}.}
\label{fig:bands}
\end{center}
\end{figure*}

Finite-size
scaling analyses~\cite{baglietto08:finite}---showing good agreement with theory for susceptibility divergence at
the phase transition---include the appearance of dense travelling bands of
particle at higher velocities
(\figRef{bands})~\cite{nagy07}.
While this could imply symmetry breaking, \figRef{bands} reveals that band orientation, as well as $\Phi$,
performs a random walk through angle space, thus not truly breaking ergodicity. Notwithstanding, the behaviour of the GTE around the phase transition at higher velocities is indeed different to lower velocities: it exhibits a peak in the long-term limit. The appearance of travelling bands coincides exactly with this noise/velocity regime, indicating that these bands could be a source of information flow.

The flat GTE exhibited in the ordered regime is a result of the approximately Gaussian heading of particles relative to their consensus vectors with variance proportional to noise. Although the continuous nature of the SVM cannot be ruled out at this stage, it seems likely that a ``discretised'' SVM would display similar behaviour. While it might seem reasonable to have investigated the behaviour of GTE in the continuous \emph{equilibrium} case for comparison, we note that the obvious contender---the classical XY model---features a BKT phase transition (at least in the 2D case)~\cite{Kosterlitz73}, which would seem to be of an entirely different nature to the transition observed in the 2D SVM model.

It is still a matter of debate whether real-world flocks are poised at criticality. Bialek~\etal~\cite{bialek12} develop a spin-wave approximation for 3-dimensional flocks of starlings, parametrised from real data. Using this model, along with analysis in~\cite{bialek14}, Bialek~\etal~discuss long-range order of the velocity (orientation) and speed fluctuations of the flock. At low noise, there is a spontaneous symmetry breaking of the continuous velocity fluctuations, leaving behind a Goldstone mode~\cite{Goldstone61}, wherein there is no energy cost for birds to perform certain changes in flight, which manifests as infinite correlation length~\cite{toner1998flocks}. Bialek~\etal~state, however, that there is no spontaneous symmetry breaking in relation to speed fluctuations, therefore no related Goldstone mode; and hence that long-range order of the flock must be a consequence of criticality.

Melfo~\cite{melfo17} instead argues qualitatively that attraction and avoidance rules common in such models do in fact spontaneously break
 continuous translational invariance, thus giving rise to a Goldstone mode and correlation lengths on the order of the system size. Melfo further states that this should appear in other examples of collective motion---particularly noting the Active-Elastic model of Huepe~\etal~\cite{Huepe15}, which achieves scale-free correlation of speed and velocity fluctuations with only attraction and avoidance interaction rules. While we cannot draw any direct conclusions here regarding criticality in real-world flocks---since the SVM utilises constant speed (and thus no speed fluctuations)---our work provides a foundation for further comparative studies of information flow as measured by GTE in alternative models that \emph{do} feature speed fluctuations.

\section{Acknowledgments}
We thank Mike Harr\'e, Joe Lizier and Guy Theroulaz for helpful discussions.

The National Computing Infrastructure (NCI) facility provided computing time for the simulations under project e004, with part funding under Australian Research Council Linkage Infrastructure grant LE140100002.

Joshua Brown would like to acknowledge the support of his Ph.D. program and this work from the Australian Government Research Training Program Scholarship.

Lionel Barnett's research is supported by the Dr. Mortimer and Theresa Sackler Foundation.

\end{document}